\documentclass[11pt,aps,preprint,pra,superscriptaddress,showkeys,amsmath,amssymb,nofootinbib]{revtex4-1}

\usepackage{graphicx}
\usepackage{dcolumn}
\usepackage{bm}
\usepackage{epsfig}
\usepackage{color}

\begin{document}

\title{Cation mono- and co-doped anatase TiO$_2$ nanotubes: An {\em ab initio} investigation of electronic and optical properties}

\author{Mohamed M.\ Fadlallah}
\email{mohamed.fadlallah@fsc.bu.edu.eg}
\affiliation{Physics Department, Faculty of Science, Benha University, Benha, Egypt}

\author{Ulrich Eckern}
\email{ulrich.eckern@physik.uni-augsburg.de}
\affiliation{Institute of Physics, University of Augsburg, 86135 Augsburg, Germany}


\begin{abstract}
The structural, electronic, and optical properties of metal (Si, Ge, Sn, and Pb) mono- and co-doped anatase TiO$_{2}$ 
nanotubes are investigated, in order to elucidate their potential for photocatalytic applications. It is found that Si 
doped TiO$_{2}$ nanotubes are more stable than those doped with Ge, Sn, or Pb.
All dopants lower the band gap, except the (Ge, Sn) co-doped structure, the decrease depending on the concentration and the type of dopant. 
Correspondingly, a redshift in the optical properties for all kinds of dopings is obtained.
Even though a Pb mono- and co-doped TiO$_{2}$ nanotube has the lowest band gap, these systems are not suitable for water splitting,
due to the location of the conduction band edges, in contrast to Si, Ge, and Sn mono-doped TiO$_{2}$ nanotubes. On the other hand,
co-doping of TiO$_{2}$ does not improve its photocatalytic properties. Our findings are consistent
with recent experiments which show an enhancement of light absorption for Si and Sn doped 
TiO$_{2}$ nanotubes.
\end{abstract}

\keywords{Nanotube, titania, metal doping, electronic and optical properties, density functional theory}

\maketitle

\section{Introduction}

Titanium dioxide (TiO$_{2}$), also known as titania, has been widely studied as a promising material for many applications
because of its low production cost, chemical stability, and non-toxicity \cite{TT,CC,FS}. 
Titania is useful for, in particular, solar cells \cite{AM}, batteries \cite{DD}, photochemical \cite{TU} and
photocatalytic \cite{LY} applications, gas sensing \cite{JZ}, and hydrogen storage \cite{MR,Gratzel2001,OK}.  
However, TiO$_{2}$ can only be activated by ultraviolet light due to its large band gap ($3.0$ eV for the rutile, 
and $3.2$ eV for the anatase phase). Therefore, engineering the band gap\footnote{The term ``band gap engineering'', introduced 
more than 30 years ago, generally refers to all attempts at modifying the band gap, e.g., by heterostructuring, combining 
suitable materials, and doping.} \cite{VN1984,Capasso1987}
of titania in order to increase its photosensitivity for visible light is a major target in photocatalyst studies. 

In recent years, various low-dimensional TiO$_{2}$ nanostructures have been prepared, such as thin films \cite{YQ}, 
nanoparticles \cite{DV,JBi}, nanowires \cite{BLiu,YJH}, and 
nanotubes \cite{PH,SPA}. TiO$_{2}$ nanotube (TNT) arrays are most interesting for applications
due to their large internal surface and highly ordered geometry \cite{GK1,GK2,MA}.  
The structural properties, stability and electronic structure of different TNT structures 
(anatase and lepidocrocite) have been discussed, e.g., in \cite{ANE}. All anatase nanotubes are semiconductors with 
direct band gaps while the lepidocrocite nanotubes are semiconductors with indirect gaps. In addition, 
anatase nanotubes were found to be most stable; their stability increases with increasing diameter \cite{AMF1,DSz,RAE}. 
The rolling of an anatase (101) sheet along the [$101$] and [$010$] directions has been used 
to build ($n$,0) and (0,$n$) TNTs, respectively \cite{AVB}.
{Further details of the geometrical properties of TNTs, in particular, about the folding procedure and the
anatase layer basic translation vectors, can be found in Refs.~\onlinecite{ANE,AMF1,DSz,RAE,AVB}}.

The experimental results show that the predominant peaks of anatase and rutile nanotubes are
(101) and (110) \cite{Varghese2003,Cesano2008}.
Recently, several mono- and co-doped TNTs have been synthesized, e.g., C \cite{Wang2018}, 
P \cite{Qin2017}, Co \cite{Yang2018}, Si \cite{Dong2018}, and Sn \cite{Li2016} mono-doped,
as well as (C/N, F) co-doped \cite{Chatzitakis2017} TNTs.
On the other hand, doped TNTs have been studied theoretically only occasionally, e.g.,
N and B doping \cite{DJM}, C, N, S, and Fe doping \cite{Piskunov2015}, (N, S) co-doping \cite{Chesnokov2017},
and nonmetal and halogen doping \cite{MMF2017}.

In the context of the present study, we note that an improvement of the photocatalytic properties of bulk TiO$_2$ has been
observed experimentally \cite{SMO2003,Ozaki2005} and calculated theoretically \cite{Yang2008,RLY} for Si doping.
Other dopings (Ge, Sn, Pb) are also known to reduce the band gap in the rutile bulk system, while Sn and Pb
doping slightly broadens the band gap in anatase TiO$_2$ \cite{RLY}. Experimentally an improvement of photocatalytic
properties was found for Sn doped bulk systems synthesized by the hydrothermal method \cite{Duan2012}. TiO$_2$
thin films doped with Si \cite{Sun2012}, Sn \cite{Arpac2007}, Pb \cite{Yu2002}, and Ge \cite{Zhou2011} have been
prepared and investigated, generally showing an improvement of photocatalytic activity upon doping.
With respect to TiO$_2$ nanotubes, a suitable doping with Si also improves the light absorption \cite{JXi,Dong2018}. 
Similar results have been found for Sn doping where, however, also a transformation from anatase to
rutile is observed \cite{Li2016}. Thus, in the light of these previous experimental and theoretical studies, and in view of
their potential high relevance for photocatalytic applications, we perform a systematic study of doping anatase TNT 
with group-IV elements, as a function of dopant concentrations; in addition, co-doping effects are also 
investigated. 
{Our focus in the present work is on the anatase nanotube, because the TiO$_2$ anatase (101) surface is known to be a
quite effective surface for solar cell applications \cite{Gratzel2009}.}

In the following, after describing the methodology (section II), we investigate the effect of the mono-dopants (Si, Ge, Sn, Pb) 
on the structure and stability of {anatase phase (8,0)} titania nanotubes (section III).
Then we study the electronic structure of doped TNTs (section IV), followed by a discussion of the optical properties (section V). 
An application of this study is the splitting of water (section VI). We close our work with a brief summary (section VII).

\section{Methodology}

We apply density functional theory (DFT) employing the generalized gradient approximation (GGA) \cite{Morales2017}
and the Perdew-Burke-Ernzerhof functional \cite{JP} as implemented in the SIESTA package \cite{JMS}.\footnote{Generally 
speaking, the accuracy of DFT-GGA calculations -- which notoriously underestimate the band gap -- is 
always an issue. In this context, we mention that this question is thoroughly discussed in a recent paper 
\cite{Morales2017}, with the conclusion that DFT-GGA is ``an empirical, yet practical'' approach. 
See also \cite{HWe,FHT,XZh} in relation to the ``scissors operation''.}
The wave functions are expanded using a local atomic orbitals basis set; the energy cutoff is $300$ Ry, 
and the Monkhorst-Pack $k$-meshes contain $1 \times 1 \times 12$ points. 
Structural relaxation is carried out with the conjugate gradient method until the net force on every atom is smaller 
than $0.04$ eV/{\AA}. As we are interested in the properties of nanotubes, a rectangular supercell, $20 \times 20 \times L$ {\AA}$^3$, 
is used,  where $L$ is the length of the nanotube along the z axis. 
The distance between two neighboring TNTs, in x and y directions, is thus $20$ {\AA}, 
which is sufficient to avoid any image interaction. 
Test calculations, changing the size of the supercell and the number of $k$-points, show the convergence of our results. 
{In particular, in order to check for the spurious dipole-dipole interaction between image
supercells, the cell has been increased to $30 \times 30 \times L$ {\AA}$^3$; however, no effect was observed. (This
aspect is discussed in detail, e.g., in \cite{Souza2013} and \cite{Rutter2019}.)} 
Spin polarized calculations have also been performed for selected systems, but no modifications were found.

\section{Optimized structure and stability}

\begin{figure}
\includegraphics[width=0.9\textwidth]{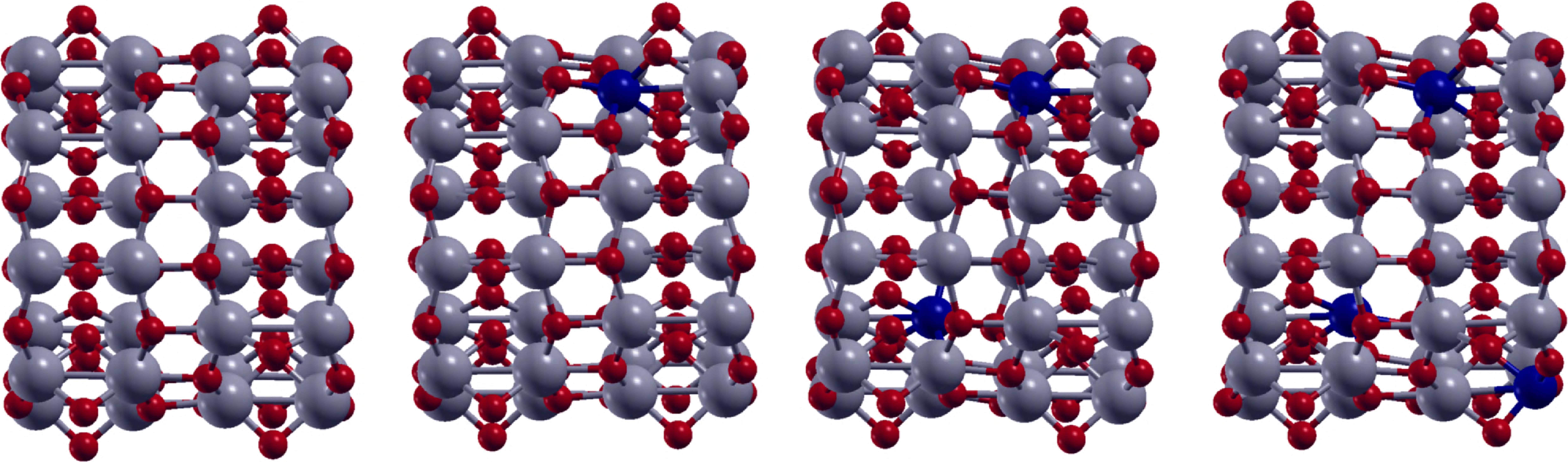}
\includegraphics[width=0.5\textwidth]{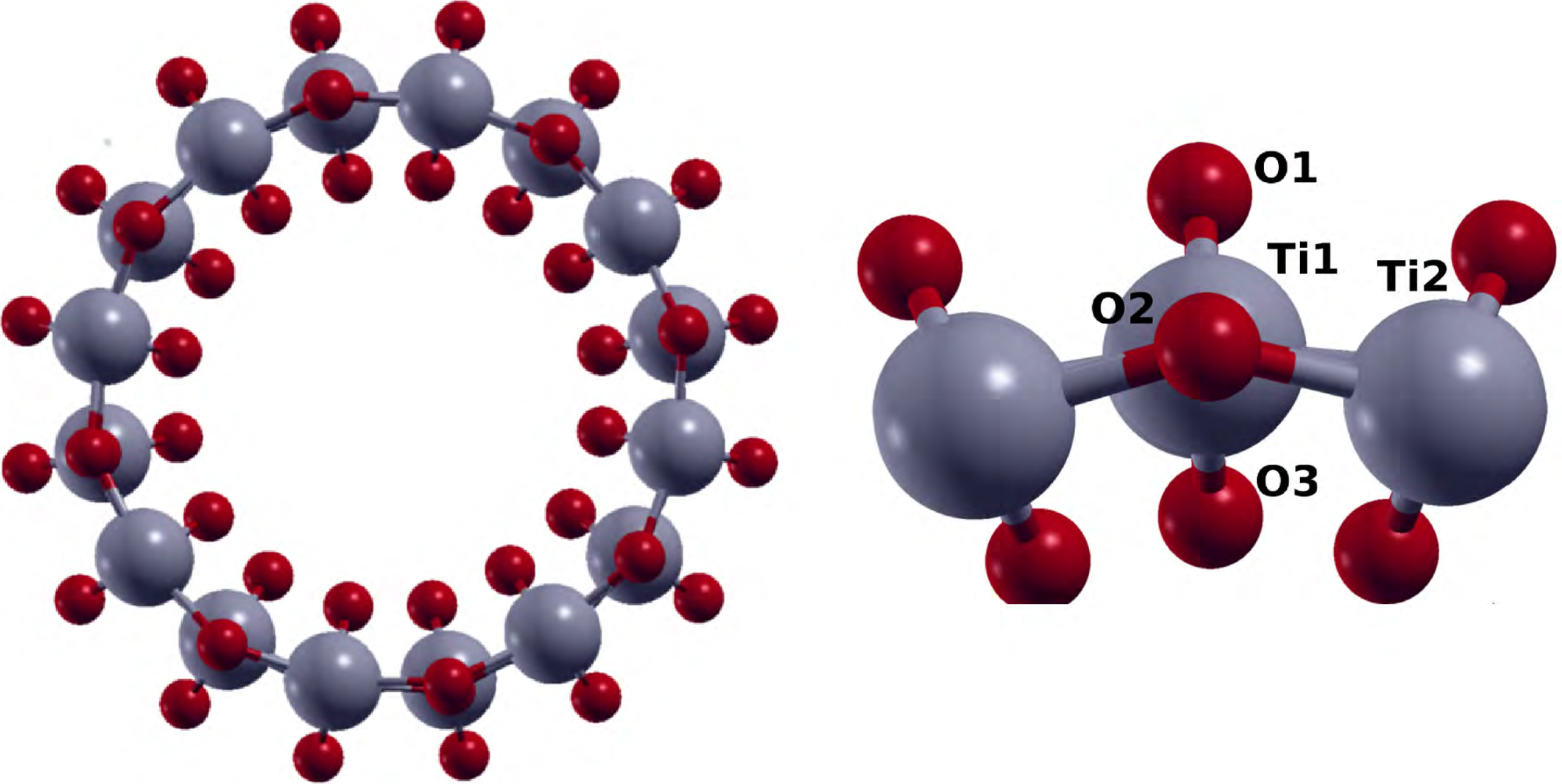}
\caption{Top part: Optimal configuration (side view) for pristine TNT, 
1\% doped TNT, 2\% doped TNT, and 3\% doped TNT (from left to right). 
Lower part: Top view of the pristine TNT (left) and detail of the wall (right).
The indicated labels are discussed in 
the main text. Red, grey, and blue spheres represent O, Ti, and dopant atoms, respectively.}
\label{fig1}
\end{figure}

The total number of the atoms in the unit cell of a TNT is related to the number of atoms in one unit cell ($48$ atoms) in the surface layer. 
We use a supercell including two TNT unit cells to study the effect of changing the doping concentration. 
Figure \ref{fig1} shows the structure of  two unit cells of (8,0) TNT. The fundamental periodic of the TNT nanotube 
(along z direction) is {found to be} $10.49$~{\AA}, which is only slightly larger than the value obtained
in a previous study, $10.13$~{\AA} \cite{RAE}.
{In that paper, only one unit cell was studied, whose length was determined to be $5.065$~{\AA}, hence we
compare with twice this value. No symmetry was preserved during structure optimization.}
Concerning other geometric parameters, we find the
inner diameter of the nanotube, cf.\ the lower part of Fig.~\ref{fig1}, i.e., between an O3 and its opposite counterpart,
to be given by $7.07$~{\AA}, while the distance between an O1 and its opposite counterpart is $12.05$ {\AA}; the diameter
with respect to the Ti atoms is $9.51$ {\AA}, and the O1--O3 distance is $2.49$~{\AA}.
The bond length Ti1--O1 (equal to Ti1--O3) is $1.84$~{\AA}, while the bond Ti2--O2 is slightly
longer, $1.95$~{\AA}, in good agreement with previous works \cite{RAE,FN}.

Cation doping of the TNT is introduced by replacing Ti atoms by the dopants. Replacing one Ti by a metal dopant corresponds 
to $\sim$ 1.0\% dopant concentration. 
If two atoms are substituted, the doping concentrations will double, and so on. These dopant concentrations are comparable
to those reported experimentally \cite{LDY}.
Though there are several possible dopant locations for 2\% and 3\% doping, we have opted in this work for configurations in 
which the dopants are as far apart as possible, namely 
{$9.8$ {\AA} for 2\%, and $9.8$ {\AA}, $9.3$ {\AA}, $7.4$ {\AA} for 3\% concentrations,} 
thereby avoiding as much 
as possible any dopant-dopant interaction. With this choice, we also avoid major distortions of the nanotube’s structure. 
Naturally, we thus exclude the possibility of dopant cluster formation (which could be an interesting question in itself
\cite{Shyichuk2016}, but is beyond the scope of the present study).

The optimized average bond lengths around the dopant atoms are listed in Tab.~\ref{table:bondlength}. 
The bond length between the dopant atom and the O atom increases as the ionic radius of the dopant increases: 
Si, Ge, Sn, and Pb, with radii $0.40$, $0.53$, $0.69$, and $1.19$ {\AA}, respectively. In comparison, the ionic 
radius of Ti$^{4+}$ is $0.61$ {\AA}. 

The charge deficiency on the metal, estimated as the difference between electronic charge densities obtained with the 
Mulliken population analysis, is also given in Tab.~\ref{table:bondlength}.
The table shows that the charge transfer from the dopant atom to the surrounding O atom is rather high for 
Si and Sn, as compared to Ge and Pb. The formation energy of doped TNTs is used to investigate the stability of the 
structures. The formation energy ($E_{\rm{form}}$) of the dopant atoms can be calculated as follows \cite{ZZ}: 

\begin{table}[htb]
\centering
{
\begin{tabular}{|c|c|c|c|c|}
\hline

Metal&Si& Ge& Sn& Pb\\  [-0.5ex]

\hline

M--O

&1.77&1.94& 2.09 &2.18    \\  [-0.5ex]

\hline

Mulliken charge

&2.12&1.48 &2.05 &1.78 \\ [1ex]

\hline

$E_{\rm{form}}$

\raisebox{2ex}
&1.5&2.1 &3.1 &4.2 \\ [1ex]

\hline

\end{tabular}
}
\caption{Bond lengths between dopant metal and oxygen, M--O ({\AA}), Mulliken charge on dopants ($e$), and formation energy, 
$E_{\rm{form}}$ (eV), for doped TNT.}
\label{table:bondlength}
\end{table}

\begin{equation}
E_{\rm{form}} = E_{\rm{M-TiO_{2}}}+\mu_{\rm{Ti}}-(E_{\rm{TiO_{2}}}+\mu_{\rm{M}}),
\end{equation}
\
where $E_{\rm{M-TiO_{2}}}$ and $E_{\rm{TiO_{2}}}$ are the total energies of the metal-doped TiO$_{2}$ and the pristine 
TiO$_{2}$ nanotube, respectively, while $\mu_{\rm{Ti}}$ and $\mu_{\rm{M}}$ denote the chemical potentials for Ti and the dopant; 
the latter are assumed to be equal to the energy of one atom in its corresponding bulk structure.

The formation energy depends on the growth conditions, which can be Ti-rich or O-rich \cite{CGV}.  
For the Ti-rich condition, thermodynamic equilibrium is assumed for the Ti bulk solid phase, thus its chemical potential 
is fixed at $\mu_{\rm{Ti}}$, while the chemical potential of O is fixed by the growth conditions. 
Under the O-rich condition, O is assumed to be in equilibrium with O$_{2}$ molecules, thus the chemical potential of O 
is $\mu_{\rm{O}}= \mu_{\rm{O_{2}}}/2$. 
We present the formation energy under the O-rich condition, which is lower than for the Ti-rich condition. 
The stability of nanotubes with dopants is in the following order: Si, Ge, Sn, Pb. 
The behavior of the formation energies can be understood, to a large extent, in terms of the dopant's 
electronegativity ({see also section VI}) given by 
$1.90$ (Si), $2.01$ (Ge), $1.96$ (Sn), and $2.33$ (Pb) {(Pauling scale)}. On the one hand, 
one notes that the formation 
energy of Si is smaller than that of the other dopants, corresponding to the fact that Si has the smallest electronegativity. 
On the other hand, the Pb formation energy is the largest, and so is its electronegativity. From this 
point of view, Ge and Sn doped TNTs are ``out of order'', which can be related to the effect of electronegativity on the 
ionic radius, implying that the formation of Sn--O bonds is more favorable than Ge--O bonds. 
This behavior of formation energies and bond lengths is very similar to the behavior of the corresponding dopant 
in bulk TiO$_{2}$ \cite{RLY}.

\section{Electronic structure}
\begin{figure}
\includegraphics[width=0.8\textwidth]{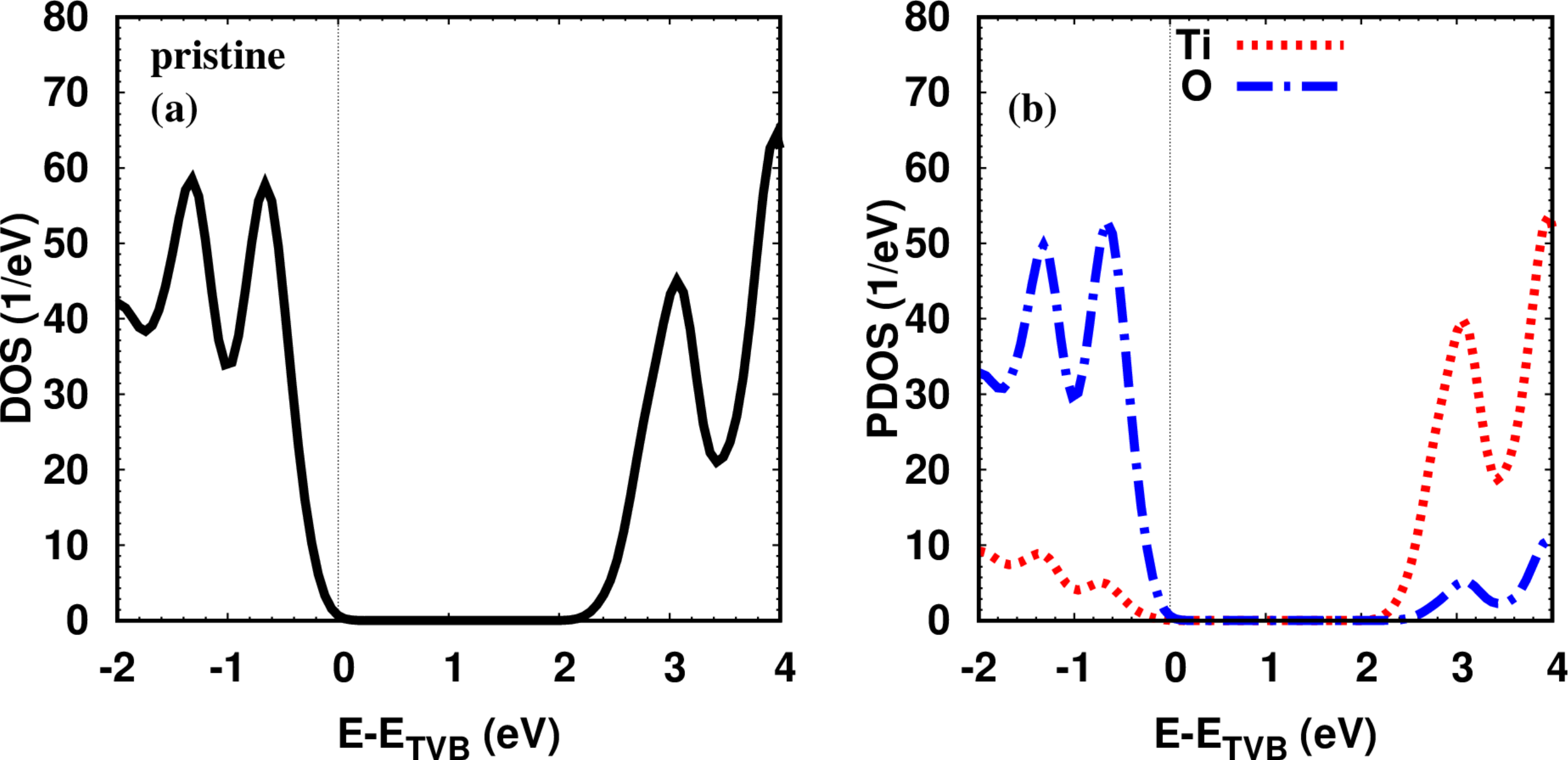}\
\caption{(a) Density of states (DOS), and (b) partial density of states (PDOS) of the 
pristine TiO$_{2}$ (8,0) nanotube. {The energy is given relative to the top of the valence band (TVB).}} 
\label{fig2}
\end{figure}

In this section, we discuss the density of states (DOS) and the partial density of states (PDOS) for the doped
TNTs under consideration, in particular, the behavior of the valence bands (VBs) and the conduction bands (CBs) upon
doping, with focus on the modifications of the energy gap. In order to present the results in a concise and systematic fashion, we
have chosen to measure the energy in the DOS and PDOS plots, Figs.~\ref{fig2}--\ref{fig6}, relative to the
top of the valence band energy, $E_{\textrm{TVB}}$, of the pristine TNT. For water splitting applications,
on the other hand, the absolute energies are required, see section VI, in particular, table \ref{table:results}.

Before going into detail, we emphasize that we have carefully checked the dependence of the 
results on the dopant positions. For example, for a concentration of 1\%, we find that the total energy for different dopant
positions varies only by less than $0.01$ eV, and no change in the DOS is obtained. For 2\% and 3\% concentrations, 
we find that the stability increases upon increasing the distance between dopants. Hence our calculations have been done 
at the largest possible distance(s) between dopant atoms.

Figure \ref{fig2}(a) shows the density of states (DOS) for pristine titania nanotubes (8,0). 
The calculated band gap is $2.20$ eV, which -- as usual in DFT-GGA -- is lower than the corresponding experimental 
gap of the TiO$_{2}$ nanotube {(3.18 -- 3.23 eV \cite{Momeni2016,Momeni2015})}. 
The Ti {\bf ($3d$)} states dominate in the unoccupied states, while the O {\bf($2p$)} states contribute mostly to 
the occupied states with 
a minor contribution to the unoccupied states, see Fig.~\ref{fig2}(b). The DOS and PDOS are 
very similar to the results obtained in \cite{FN}.

\subsection{Mono-doped TNTs}

\begin{figure}
\includegraphics[width=0.9\textwidth]{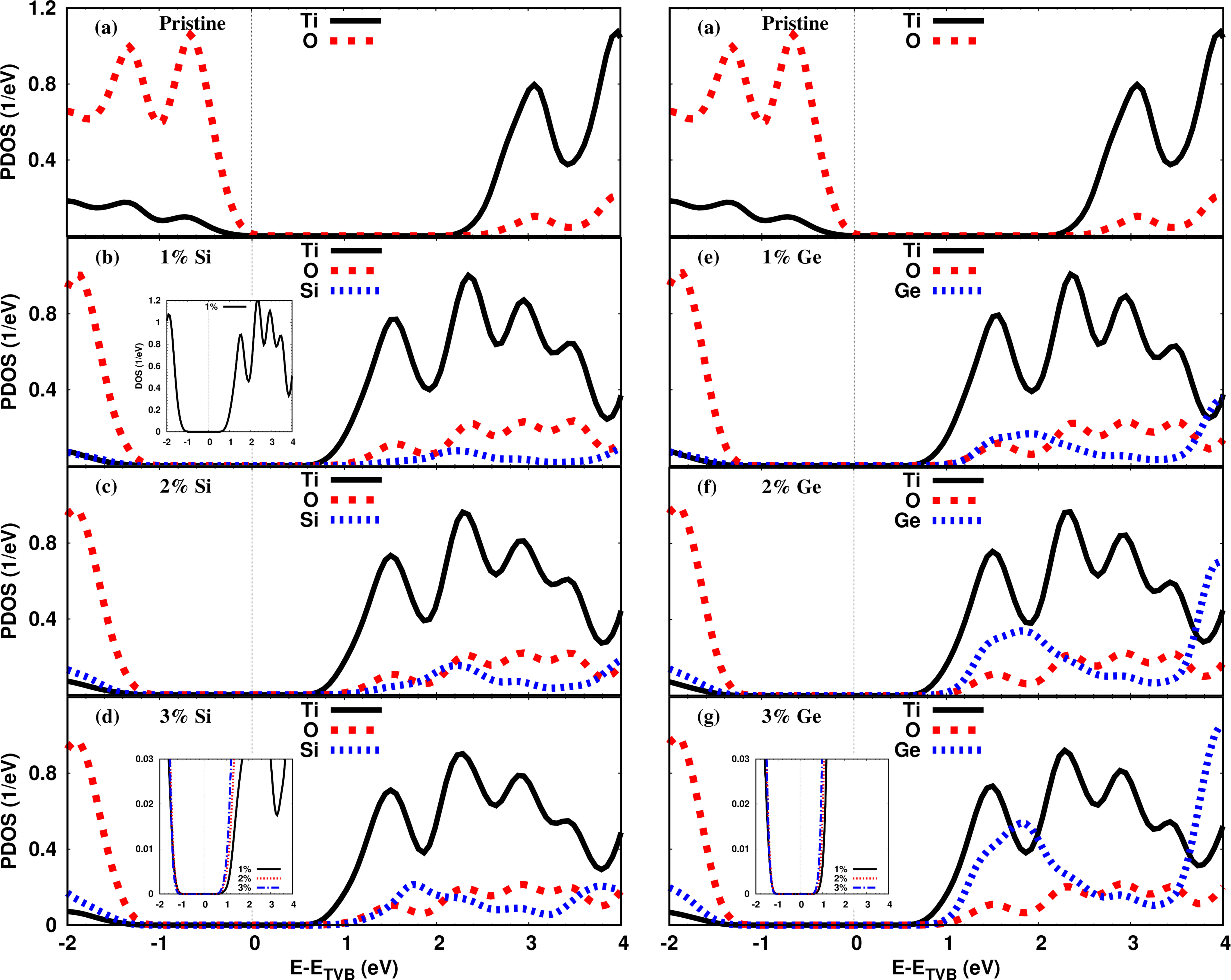}
\caption{Partial density of states (PDOS) for (a) pristine TNT (shown at the top of both columns), and 
mono-doping at different concentrations: for (b) 1\%, (c) 2\%, and (d) 3\% Si,
and for (e) 1\%, (f) 2\%, and (g) 3\% Ge. (Ti, O) states are scaled down by a factor 50 to allow easy comparison. 
The energy is given relative to the top of the valence band (TVB) of the pristine TNT. The inset in (b), (d), and (g) shows 
the corresponding total DOS (Si doping), and PDOS (Si, Ge) on an even more reduced scale.} 
\label{fig3}
\end{figure}

{\em Si-doping.}
The effect of Si doping at different concentrations on the electronic structure of TNT is shown in Fig.~\ref{fig3}(b-d). 
The band gap is $1.80$ eV for 1\% doping, less by 0.40 eV than that of the pristine TNT. The corresponding total DOS is similar 
to the pristine DOS, however, with a smaller band gap, see inset of Fig.~\ref{fig3}(b) as compared to Fig.~\ref{fig2}(a).
When the concentration increases to 2\% and 3\%, 
we find that the Si--Si distance decreases to 9.4 {\AA} after optimization for 2\% concentration, 
and to $9.4$ {\AA}, $9.0$ {\AA}, and 7.3 {\AA} between different pairs of Si atoms for 3\%. These values have to be
compared with the original Ti--Ti distance of $9.8$ {\AA} for 2\%, and $9.8$ {\AA}, $9.3$ {\AA}, and $7.4$ {\AA} for 
3\% concentrations. The band gap remains at $1.80$ eV for 2\%, and increases to $1.86$ eV for 3\%. 
The computed band gap reduction for the corresponding doped bulk system
is found to be slightly smaller, 0.20 eV \cite{RLY}. (In that paper, only 2\% doping was studied.)

Concerning the detailed behavior, we note that on the scale of the figure an almost rigid,
concentration independent downshift of about $1.3$ eV of the VB is observed, accompanied
by a slight ``smearing'' of the oscillations which are visible below $-0.5$ eV in the pristine
PDOS. The PDOS  shows that the dopant states start contributing above $1$ eV, with a distinct maximum
at about $2.2$ eV. The dopant contribution is rather small, but increases continuously with
increasing concentration. Comparing with the Ti PDOS, Fig.~\ref{fig3}(a), we realize that while the
onset of Si states is clearly lower than the onset of the pristine Ti states, the latter coincides with the
maximum of the Si PDOS. The evolution of the maximum of the Si PDOS can be seen more
clearly in Fig.~\ref{fig3}(b-d). It is located near $2.2$ eV for 1\% and
2\% concentration, Fig.~\ref{fig3}(b,c), then shifts downwards to about $1.8$ eV for 3\% concentration,
see Fig.~\ref{fig3}(d). As the location of dopant states shifts closer to the CB edge, the band gap increases.
The decrease of the band gap -- as compared to the pristine TNT -- is consistent with the observed increase of optical
absorption of TNT upon Si doping \cite{Dong2018,YSu,JXi}.

{\em Ge-doping.}
The optimized Ge--Ge distances are slightly larger than the Si--Si values, consistent with the increase in ionic radius,
namely 9.7 {\AA} for 2\% concentration (and hence only 0.1 {\AA} smaller than the original Ti--Ti distance),
and 9.7 {\AA}, 9.2 {\AA}, and 7.5 {\AA} for 3\% concentration. Figure \ref{fig3}(e,g) shows that the band gap is $1.86$ eV, 
at any concentration, which is less than the pristine band gap but larger than that for Si doping TNT at 1\% and 2\% concentration, 
because the location of Ge states is closer to CB edge than the Si states at these concentrations.
The band gap does not depend on the concentration because the dopant states peak position ($1.86$ eV) is rigid (Fig.~\ref{fig3}(d)). 
The peak in the Ge PDOS can be attributed to the fact that the Ge ionic radius and electronegativity are only slightly
different in comparison to Ti.
The shifts of the VB and the CB are quite similar to the case of Si doping, even though the Ge states give a stronger 
contribution in the PDOS, Fig.~\ref{fig3}(d). Accordingly, the DOS of the Ge doped system at different concentrations 
is very similar to the DOS of the Si doped structures, see inset of Fig.~\ref{fig3}(b)). 
Again, the bulk gap reduction was reported to be slightly smaller, only $0.15$ eV \cite{RLY} compared to 
the present $0.34$ eV.   

\begin{figure}
\includegraphics[width=0.9\textwidth]{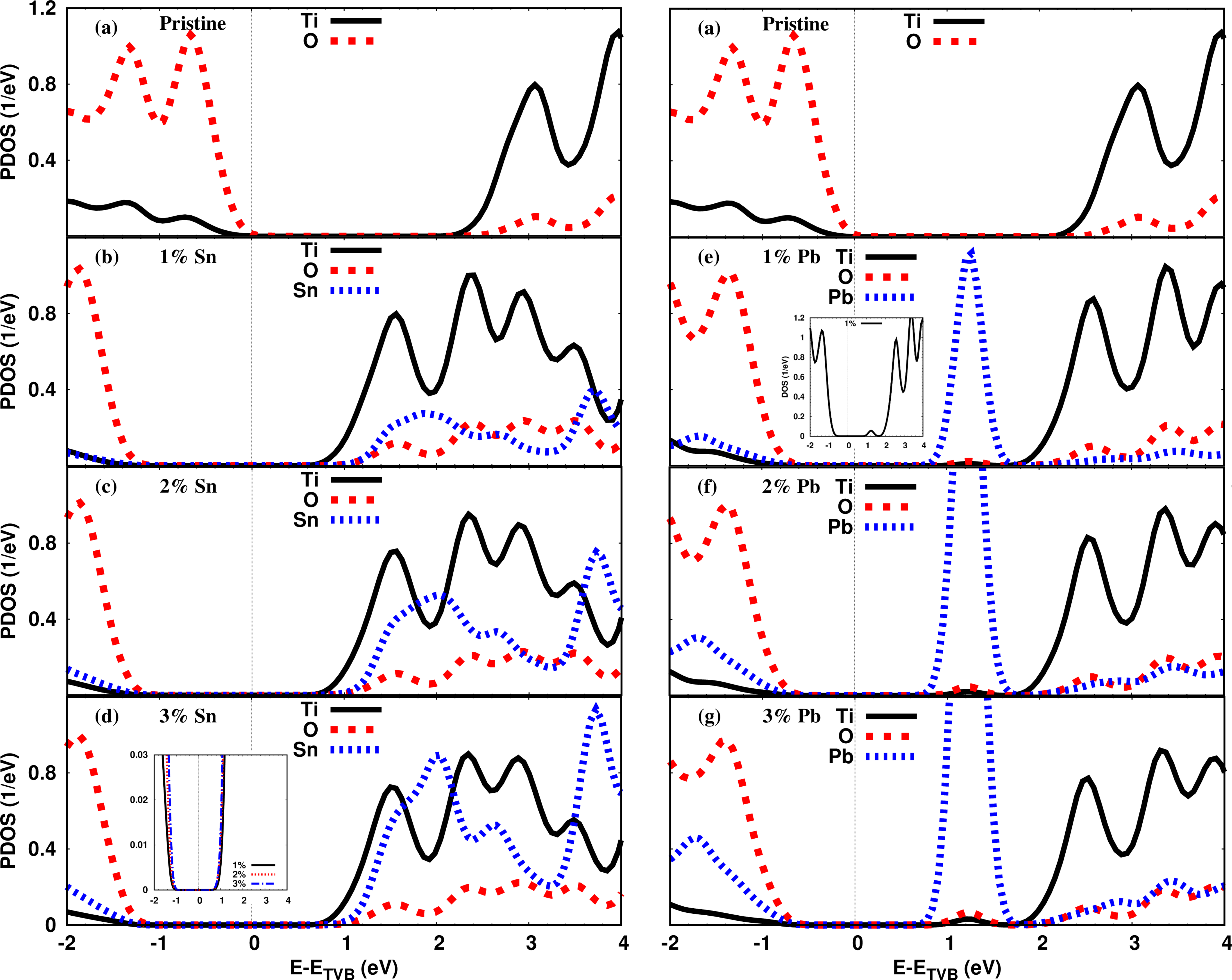}
\caption {Partial density of states (PDOS) for (a) pristine TNT (shown at the top of both columns), and mono-doping 
at different concentrations: for (b) 1\%,  (c) 2\%, and (d) 3\% Sn,
and for (e) 1\%, (f) 2\%, and (g) 3\% Pb.  (Ti, O) states are scaled down by a factor 50 to allow easy comparison. 
The energy is given relative to the top of the valence band (TVB) of the pristine TNT. The insets in (d) and (e) show 
the Sn PDOS on an even more reduced scale, as well as the total DOS (Pb doping).}
\label{fig4}
\end{figure}

{\em Sn-doping.}
For Sn doped TNTs, we find that for 2\% concentration the Sn--Sn distance is 9.8 {\AA} which is the same as the original
distance of Ti atoms. Regarding 3\% concentration, the distances are 9.8 {\AA}, 9.6 {\AA}, and 7.6 {\AA} 
which are larger than original distances. Figure~\ref{fig4}(b) shows that the contribution of Sn states at 1\% concentration 
is similar to Ge doping structure at 2\% in the energy range $-2.0$ eV to $2.5$ eV. 
The distinct peak appears at $1.8$ eV which is the same as for Si at 3\% and Ge at any concentration. 
Hence the band gap also is $1.86$ eV. When the Sn concentration increases, the distinct peak sightly shifts to 
higher energy (Fig.~\ref{fig4}(b,c)), which is in the opposite direction compared to Si doping with increasing concentration. 
The distinct peak is located at $2.0$ eV, and we find the gap to be $1.83$ eV. Due to the similarity of the Ti PDOS of doped 
Sn structures with the corresponding one of Si doped structures, the general behavior of the DOS for Sn doping is similar
to the DOS for Si doping. The computed reduction of the band gap through Sn doping also is 
in good agreement with the corresponding light absorption experiment \cite{Li2016}.

{\em Pb-doping.}
Within the mono-doped series, we finally consider Pb. The optimized Pb--Pb distances are 10.0 {\AA} for 2\% concentration, 
and 10.0 {\AA}, 9.6 {\AA}, and 7.7 {\AA} for 3\% concentration. These values are larger than the original distances of the 
host atoms. The distinct peak of Pb states is not only located at a lower energy ($1.2$ eV) as compared to the peak of the 
previously discussed dopants, but also clearly lower than the Ti CB states, see Fig.~\ref{fig4}(e-g), such that a separate
dopant peak appears in the corresponding DOS. These states decrease the band gap to $1.56$ eV, which is the lowest band gap 
in comparison to the other dopants at any concentration. On the other hand, the shift of the VB and CB edges is $0.5$ eV and $0.3$ eV, 
respectively, downwards in energy which is less than the corresponding values for the other systems. For the resulting DOS
see inset of Fig.~\ref{fig4}(e), as compared to the inset of Fig.~\ref{fig3}(b).
Increasing the concentration of Pb, the band gap slightly decreases to $1.50$ eV and $1.44$ eV, for 2\% and 3\%, respectively. 
The PDOS shows that the majority of the additional states derive from the Pb states, see Fig.~\ref{fig4}(f-g). 
The peak energy of the Pb states hardly changes with increasing doping.

In order to obtain a better understanding of the systematics of the above results, and of those presented in the
following subsection, we emphasize that the relevant aspect is the energetic location of cation dopant states
relative to the conduction band, in which the Ti $3d$ states dominate. In particular, we have been able to relate
the characteristic concentration and dopant dependent shifts of the dopant PDOS, especially
the distinct peak, to the behavior of the energy gap.
In comparison to the doped bulk system \cite{RLY}, we note first of all, that the gap reduction for the doped TNT
generally is stronger than in the bulk, where, in fact, a gap enhancement was found for 2\% Sn or Pb doping. However,
the stronger effect of doping -- compared to the bulk system -- appears reasonable since the disturbances created by
dopants are expected to have a stronger influence in a reduced-dimensionality system like a nanotube. With respect to
the location of dopant states, we note that their respective peak energies obey the following inequality:
$E_{\textrm{Pb(6s)}} < E_{\textrm{Ti(3d)}} < E_{\textrm{Sn(5s)}} < E_{\textrm{Ge(4s)}} < E_{\textrm{Si(3s)}}$. In
particular, the Si, Ge, and Sn states are well within the conduction band, such they are not able to form distinct
dopoant states below the CB. Instead, they ``only'' reduce the energy gap. However, there is
no obvious trend -- except for the relation to the distinct dopant PDOS peak, see above -- when the dopant 
concentration is increased, see table~\ref{table2}: When the Si concentration
is increased, the gap slightly increases, which likely can be related to the fact that a rather large geometric
disturbance is created by Si which has the smallest ionic radius; and that this disturbance is reduced upon doping,
at least for 3\%. While the gap for Ge doping is concentration independent, it is found to slightly decrease for Sn
doping, which is reasonable since the ionic radius of Sn is larger than the ionic radius of Ti. However, these are
rather subtle effects, and we believe it is hardly possible to identify a single ``cause'' for the systematics.

The only clear-cut case in the considered series is Pb, where the dopant states are strong and located in energy
clearly below the conduction band. Thus a separate peak in the DOS is formed, whose amplitude increases with dopant
concentration. In addition, a strong reduction of the energy gap is found.

\subsection{Co-doped TNTs}

\begin{figure}
\includegraphics[width=0.9\textwidth]{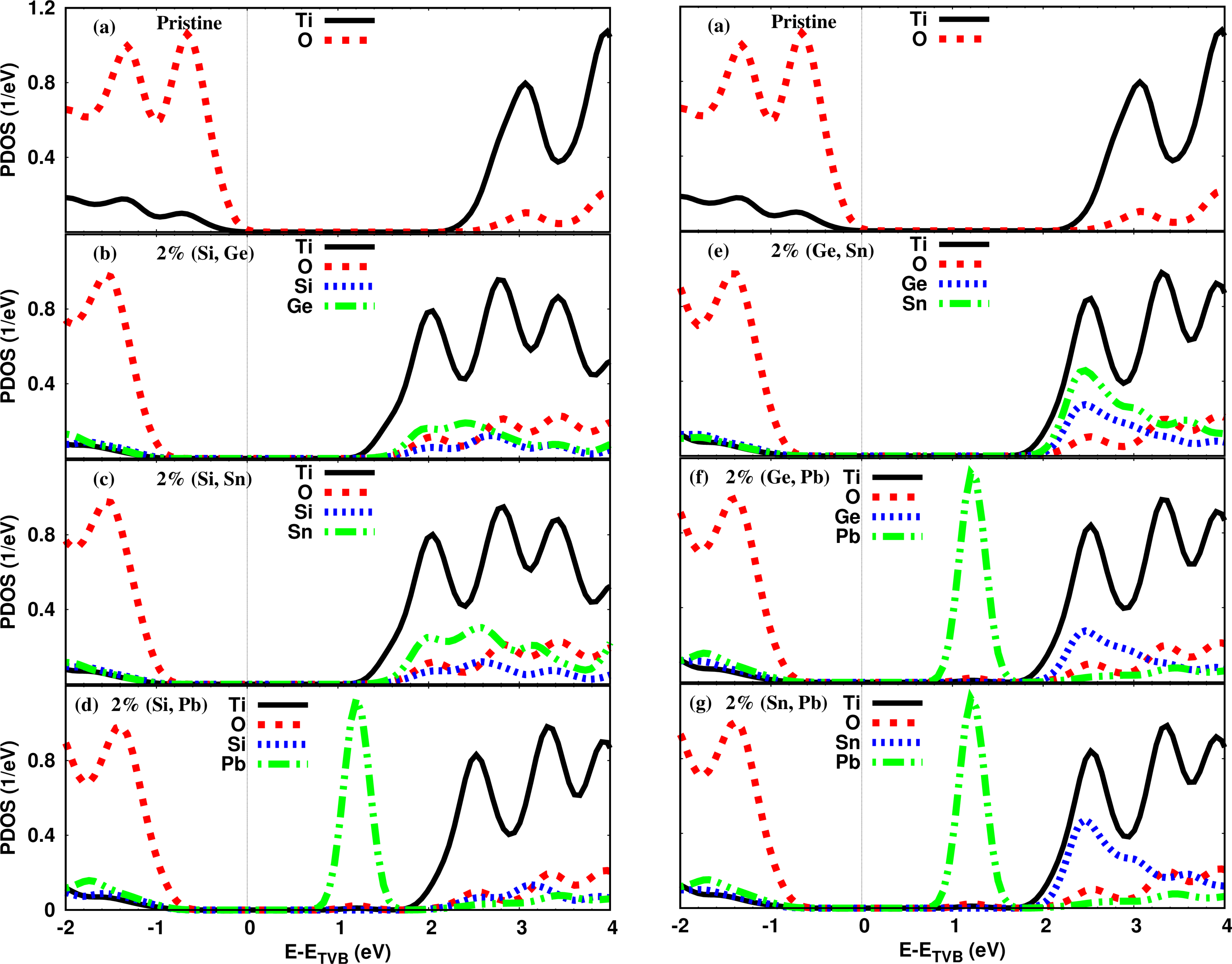}
\caption {Partial density of states (PDOS) for (a) pristine TNT, and 2\% co-doping: (b) (Si, Ge), (c) (Si, Sn), (d) (Si, Pb), 
(e) (Ge, Sn), (f) (Ge, Pb), (g) (Sn, Pb). The (Ti, O) states are scaled down by a factor 50 to allow comparison. The energy is
given relative to the top of the valence band (TVB) of the pristine TNT.}
\label{fig5}
\end{figure}

Turning finally to co-doped TNTs, the doping concentration is 2\% for two different substitutional atoms, and 3\% for 
two atoms from the same kind plus one doping atom from another kind. We first study the effect of co-doping at 2\%. Starting with
Si--Ge, the optimized distance is 9.5 {\AA}. This is approximately the average of the Si--Si and Ge--Ge distances at 2\% concentration. 
Figure~\ref{fig5}(b) shows the effect of (Si, Ge) co-doping on the electronic structure. The overlap between Si, Ge, and Ti states 
appears near the CB band, therefore the bands shift to lower energy.
This shift is less than the corresponding one in the case of 1\% and 2\% mono-doping with Si and Ge by $0.6$ eV. The
(Si, Ge) co-doped TNT has a band gap of $1.92$ eV, larger than the gap for Si and Ge mono-doping at any concentration. 
As we go down the group of dopants in the periodic table (4A), the dopant-dopant distance increases to 9.6 {\AA} for 
Si--Sn and 9.7 {\AA} for Si--Pb. The overlap between the dopant and the host atom states also increases slightly near 
the CB edge, so the band gap of (Si, Sn) is $1.98$ eV, see Fig.~\ref{fig5}(c). Figure~\ref{fig5}(d) shows the PDOS of 
the (Si, Pb) co-doped system, which is very similar to the DOS of the Pb mono-doped TNT. It is characterized by 
Pb dopant states below the conduction band. The band gap for this co-doping is $1.50$ eV, smaller than
the gap of the 1\% and equal to the 2\% mono-doped Pb system. 

As compared to the co-doped structures discussed above, the bands of (Ge, Sn) shift to higher energy, and the band gap increases
to $2.3$ eV, Fig.~\ref{fig5}(e), clearly larger than the gap of pristine TNT. The increase in the band gap can be attributed 
to the strong interaction (bonding) between the dopant (such as Sn) and the Ti CB states in the energy range 1.8--2.5 eV, 
see Fig.~\ref{fig5}(e). 
The distance between Ge--Sn is similar to the Si--Sn distance (9.6 {\AA}). If Sn is replaced by Pb, the distance of 
Ge--Pb increases to 9.8 {\AA}, which is less than the Pb--Pb distance at 2\%. Due to the interaction between Ge and Ti states 
near the CB edge, the Pb states slightly move towards higher energy, so the band gap slightly increases to $1.56$ eV as 
compared to (Si, Pb), see Fig.~\ref{fig5}(f). The last 2\% co-doped system is (Sn, Pb), Fig.~\ref{fig5}(g), 
with 9.9 {\AA} Sn--Pb distance. The distinct peak of the Sn dopant 
is located at the same position as for Ge ($2.4$ eV), Fig.~\ref{fig5}(c), and the Pb midgap states 
remain in their place, thus the band gap does not change. 
The band gap of co-doped systems at 2\% concentration is larger than the band gap of the
individual corresponding mono-doped structures because of a good co-dopant states interactions near the CB edge, 
except for Pb doping. Table~\ref{table2} summarizes the band gap values of all structures considered.

\begin{figure}
\includegraphics[width=0.9\textwidth]{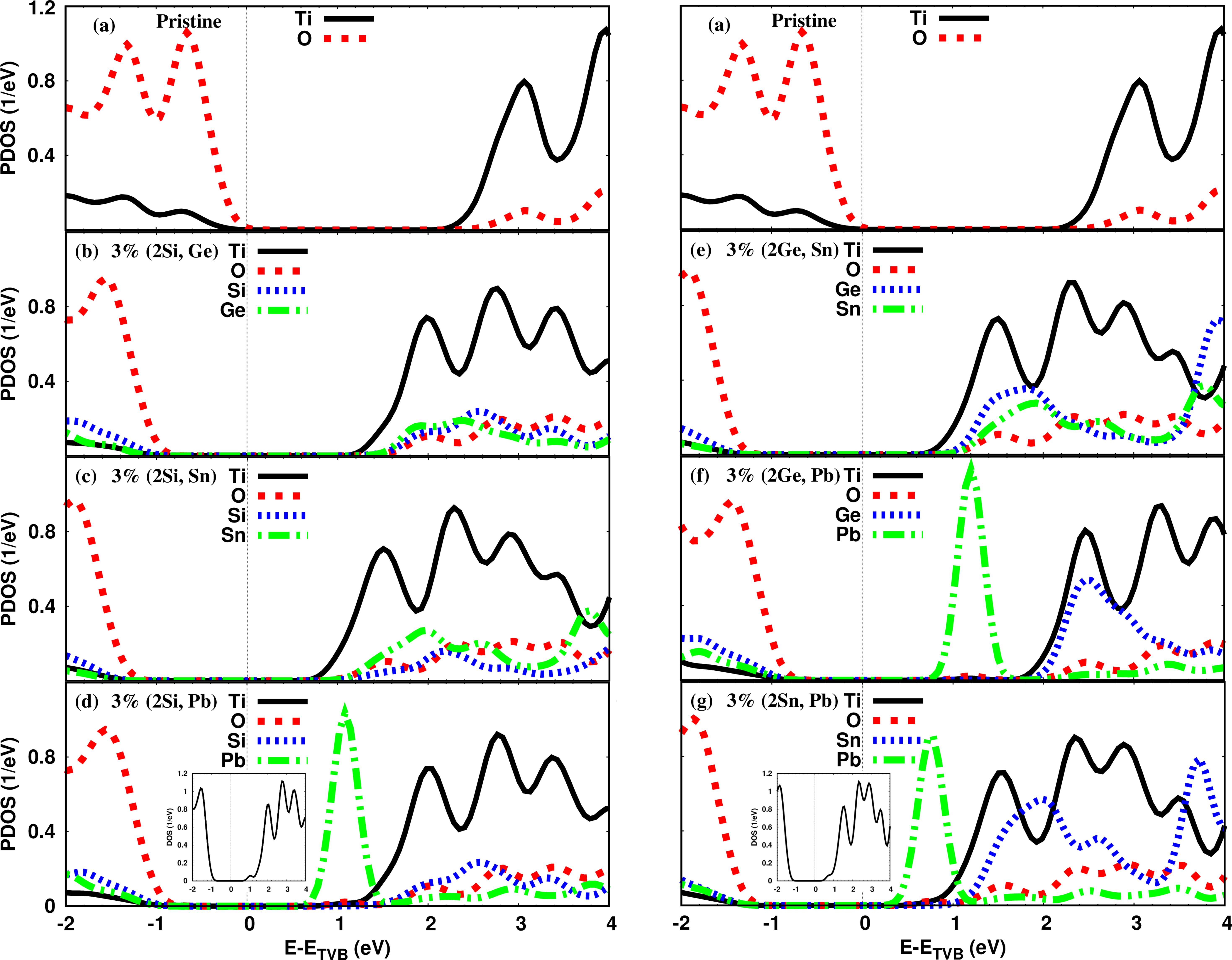}
\caption {Partial density of states (PDOS) for (a) pristine TNT, and for 3\% co-doping: (b) (2Si, Ge), (c) (2Si, Sn), (d) (2Si, Pb)
(e) (2Ge, Sn), (f) (2Ge, Pb), and (g) (2Sn, Pb). The (Ti, O) states are scaled down by a factor 50 to allow easy comparison. The energy is
given relative to the top of the valence band (TVB) of the pristine TNT. The insets in (d) and (g) show the corresponding total DOS.}
\label{fig6}
\end{figure}

Last we study co-doped TNTs at high concentration, i.e., 3\%. The distances in the (2Si, Ge) co-doped structure 
are 9.1 {\AA}, 7.3 {\AA} for Si--Ge, and 9.4 {\AA} for Si--Si, and for (Si, 2Ge) are 9.1 {\AA}, 7.4 {\AA} for Ge--Ge 
and 9.6 {\AA} for Ge--Si. 
Figure ~\ref{fig6}(b) shows the PDOS of (2Si, Ge) which is practically identical to the (Si, Ge) case. 
Also there is no change when another configuration, (Si, 2Ge), is considered. The electronic structures of 
(Si, Ge)/(2Si, Ge)/(Si, 2Ge) co-dopants do not depend on the concentration of the individual dopants 
because all configurations have a similar effect at the same energy. Regarding (2Si, Sn), we find, see Fig.~\ref{fig6}(c), 
that the overlap between states in the CB reduces the band gap as compared to (Si, Sn) by $\sim$ $0.3$ eV, 
and the gap becomes $1.86$ eV, less than the band gap of the same co-doped system at 2\% concentration. 
The DOS of the (Si, 2Sn) system is practically the same as the (2Si, Sn) DOS, even though the distances
differ slightly: 9.1 {\AA}, 7.5 {\AA} for Si--Sn, and 9.8 {\AA} for Si--Si, for the former, and 
9.2 {\AA}, 7.4 {\AA} for Si--Sn, and 9.4 {\AA} for Si--Si for the latter case.

For (2Si, Pb) co-doping, Fig.~\ref{fig6}(d), the PDOSs show that the CB and Pb midgap states shift towards
lower energy by $0.6$ eV and $0.2$ eV, respectively, as compared to the same co-doped system at low concentration, 
Fig.~\ref{fig5}(d). This relatively strong shift in the CB is due to the shift of the corresponding Si states.
The overlap between CB and Pb states is most pronounced at $1.4$ eV, which results in a small shoulder 
in the DOS (see inset of figure). 
This reduces the band gap to $1.44$ eV, less than the corresponding one for 2\% co-doping but equal 
to the band gap of 3\% Pb mono-doping. The PDOSs of (Si, 2Pb) show the same gap and DOS shape as (Si, Pb) due to
the dominant effect of the Pb states.

For (2Ge, Sn)/(Ge, 2Sn) co-doping, Fig.~\ref{fig6}(e), the PDOSs are very similar to the case of (2Si, Sn), Fig.~\ref{fig6}(c), 
with a small shift of bands to higher energy. The band gap is $1.86$ eV, the same as for Ge mono-doping at any concentration. 
The PDOSs of (2Ge, Pb)/(Ge, 2Pb) are the same as for (Ge, Pb), with the same band gap. 
The last 3\% co-doped structure is (2Sn, Pb): as compared to the (2Si, Pb) system, the PDOS is very similar
with respect to the Pb contribution, but there is a shift in energy due to the Sn states (in comparison with the
Si states), consistent with what we observed for the case of Sn versus Si mono-doping.
As compared to (Sn, Pb) co-doping, the CB and Pb states for (2Sn, Pb) are lower in energy due to the higher 
concentration of Sn; the band gap is $1.50$ eV. For (Sn, 2Pb), due to the high concentration of Pb, the PDOS is similar 
to (Sn, Pb). The band gaps of the 3\% co-doped structures are also presented in table \ref{table2}.

We note that for a given co-doped system, say, (2X, Y), there are different possibilities to position the X and Y atoms. 
We have considered such different cases, and we have confirmed that the DOSs are not affected.

\begin{table}
\centering
{
\begin{tabular}{|c|c|| c|c|c||  c|c|c||  c|c|c|| c|c|c|| c|c|c|}
\hline 
 & pr &\multicolumn{3}{c|}{Si}&\multicolumn{3}{c|}{Ge}&\multicolumn{3}{c|}{Sn}&\multicolumn{3}{c|}{Pb}&\multicolumn{3}{c|}{Si/Ge}\\\hline
 &   &  $1$\% &  $2$\%&  $3$\%  &  $1$\% &  $2$\%&  $3$\% &  $1$\%  &  $2$\%&  $3$\%   &  $1$\% &  $2$\%&  $3$\% &  (1,1) & (2,1)  &(1,2) \\\hline
$E_{\rm gap}$&2.20 & 1.80 & 1.80 & 1.86  & 1.86&1.86& 1.86 & 1.86& 1.83&1.83 &1.56&1.50 &1.44 & 1.92& 1.92& 1.92 \\\hline \hline
 &pr &\multicolumn{3}{c|}{Si/Sn}&\multicolumn{3}{c|}{Si/Pb}&\multicolumn{3}{c|}{Ge/Sn}&\multicolumn{3}{c|}{Ge/Pb}&\multicolumn{3}{c|}{Sn/Pb}\\\hline
 &  &(1,1) & (2,1)  &(1,2) &(1,1) & (2,1)  &(1,2)  & (1,1) & (2,1)  &(1,2)  & (1,1) & (2,1) &(1,2) & (1,1) & (2,1) &(1,2) \\\hline        
$E_{\rm gap}$&2.20 &1.98&1.86& 1.86  & 1.50&1.44& 1.50 & 2.34& 1.86&1.86 &1.56&1.56 &1.56  &1.56&1.50&1.56 \\\hline
\end{tabular}
}
\caption{Calculated band gap values (eV) for all concentrations and configurations considered; ``pr'' denotes the pristine TNT.}
\label{table2}
\end{table}

\section{Optical properties}

The optical properties of a semiconductor photocatalyst are closely
related to its electronic structure. The decrease of the band gap for all mono-dopants as compared to pristine TNT, see Fig.~\ref{fig3}, leads to a redshift of
the optical absorption edge. This redshift depends on the kind of dopant and the concentration. 
Clearly, several factors are relevant for the differences between doped bulk TiO$_{2}$ and doped TNTs, namely the geometry, 
electronic structure, and the interaction between dopant and neighboring Ti and O atoms, with the general tendency 
of reducing the optical gap. As is apparent from Fig.~\ref{fig4}, this leads to a shift of the absorption edge towards higher 
wavelengths, most pronounced for Pb mono- and (2Pb, Ge) co-doping. In contrast, a reduction of the optical gap upon 
doping in the {\em bulk} system is only found for Si and Ge doping \cite{RLY}. Our results agree qualitatively with the recently 
observed gap reduction for Sn doped TNTs \cite{Li2016}.

The optical absorption is related to the complex dielectric function $\varepsilon(\omega)=\varepsilon_{1}(\omega)+i\varepsilon_{2}(\omega)$, 
with $\omega$ the frequency. The imaginary part is calculated from the momentum matrix elements 
between the occupied and unoccupied states, and the real part subsequently from the Kramers-Kronig relation. 
The absorption coefficient then is given by \cite{C4}
\begin{equation}
\alpha(\omega) = \sqrt{2}\omega\sqrt{\sqrt{\varepsilon_{1}^{2}(\omega)+\varepsilon_{2}^{2}(\omega)}-\varepsilon_{1}(\omega)}.
\end{equation}
A ``scissors operation'' \cite{HWe,FHT,XZh} of 1.0 eV, which corresponds to the difference between the calculated 
and the experimental gap ($3.2$ eV) for pristine TNT, is also used for the doped system.

{As a side-remark, we wish to add that this ``operation'' -- adding {\it ad hoc} a correction $\Delta$ to the conduction
band energies, such that the calculated energy gap plus $\Delta$ (here 1.0 eV) equals the experimental gap, is pretty much
standard. It relates to the well known problem of density functional theory that the gap calculated from the Kohn-Sham
orbitals almost always is by far too small \cite{Wang1981, Godby1988}. Another way out of this problem is to extend DFT
and include so-called $GW$ corrections, see \cite{Morales2017} and references therein; but this approach is computationally
quite costly, and hence not practical for systematic studies of doped systems.}

\begin{figure}
\includegraphics[width=0.45\textwidth]{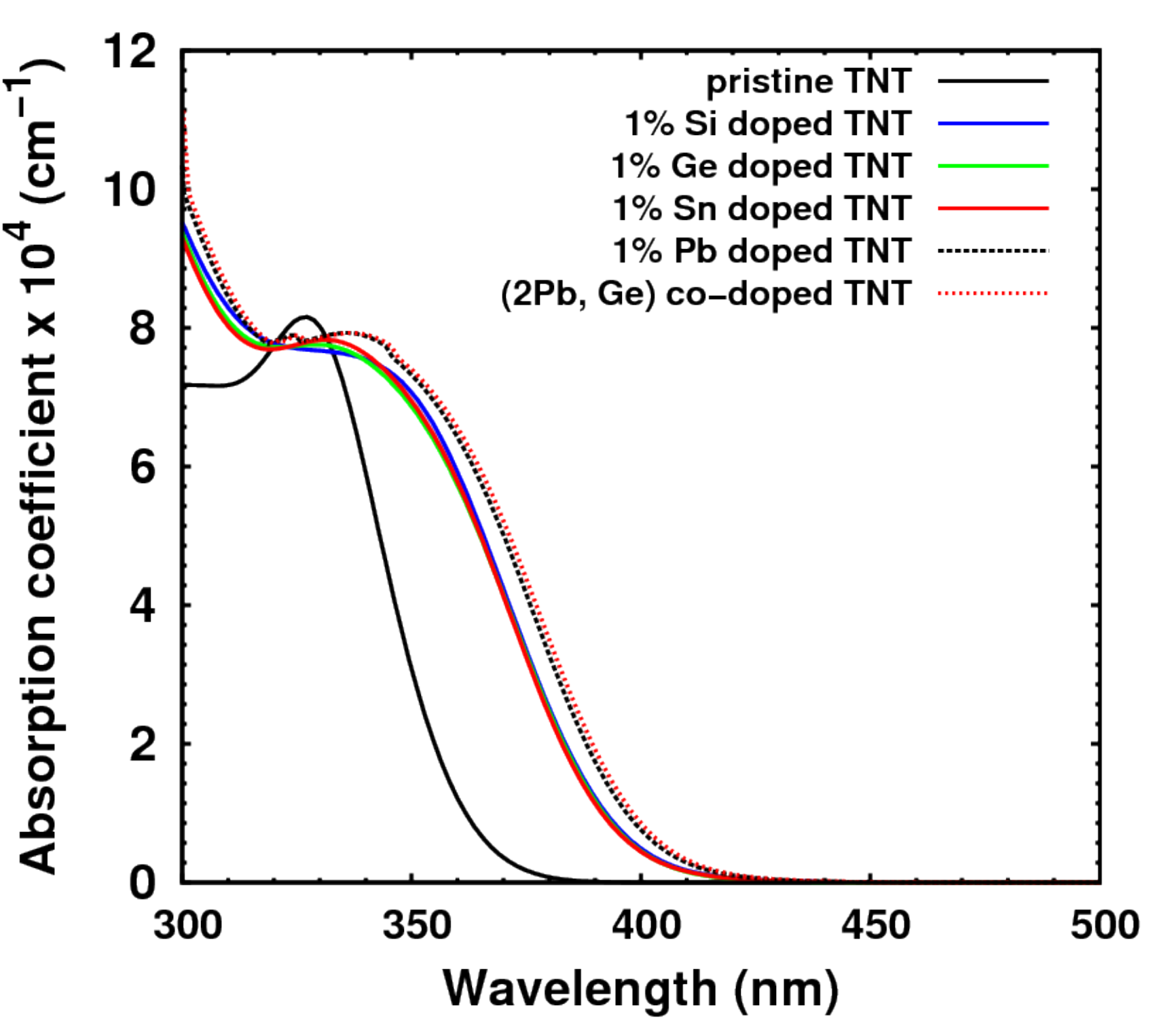}\
\caption{Absorption coefficients of pristine, mono- and co-doped TiO$_{2}$ nanotubes.} 
\label{fig7}
\end{figure}

A pristine TiO$_{2}$ nanotube can only absorb the narrow UV light ($370$ nm), but shows no absorption for visible light, 
see Fig.~\ref{fig7}. The calculated optical absorption spectra for all mono-doped TNTs
show absorption in the visible-light region, namely in the range $380-410$ nm. Also, a redshift is 
apparent for all mono-doped TNTs, consistent with the earlier discussion.

\section{Application: Water splitting}
The improvement of the visible light activity of TiO$_{2}$ is very important for water splitting (H production) \cite{Chen2010,Modak2015}.
{In this context, it is important to note that the absolute values of the conduction and valence band edges,
$E_{\textrm{CBE}}$ and $E_{\textrm{VBE}}$, are required. This issue has been discussed extensively in the literature; see, e.g.,
Ref.~\onlinecite{Gratzel2001}, or, more recently, Ref.~\onlinecite{Wang2016}. In short \cite{MMF2017}, the conduction band edge is computed from
the empirical relation $E_{\textrm{CBE}} = \bar X - 0.5 E_{\textrm{gap}} - 4.5$ eV, where $\bar X$ denotes the geometric mean of the 
electronegativities of the constituents (e.g., $\bar X =(\chi_{\textrm{Ti}}\chi_{\textrm{O}}^{2})^{1/3} = 5.80$ eV for the pristine case,
using experimental values \cite{Putz2005,ECM,DAV}), and $E_{\textrm{gap}}$ is the scissors-corrected energy gap. 
Then $E_{\textrm{VBE}} = E_{\textrm{CBE}} + E_{\textrm{gap}}$.
Calculations of the conduction band edge (CBE) and the valence band edge (VBE) have shown that the 
CBE of anatase TiO$_{2}$ is located at $-0.29$ eV, and the VBE at $2.91$ eV \cite{XYu}.
Note that these band edges are measured with respect to the normal hydrogen electrode (NHE) potential of the reduction and 
oxidation levels of water: the reduction level (H$^{+}$/H$_{2}$) is located at $0$ eV, and the oxidation 
level (H$_{2}$O/O$_{2}$) at $1.23$ eV, respectively. Thus the CBE is ``above'' the water reduction (H$^{+}$/H$_{2}$) level,
and the VBE ``below'' the water oxidation (H$_{2}$O/O$_{2}$) level, in the standard representation \cite{Gratzel2001}.
}

\begin{table}
\centering
{
\begin{tabular}{|c|c|| c|c|c|| c|c|c|| c|c|c|| c|c|c|}
\hline
 &pristine &\multicolumn{3}{c|}{Si dopant}&\multicolumn{3}{c|}{Ge dopant}&\multicolumn{3}{c|}{Sn dopant}&\multicolumn{3}{c|}{Pb dopant} \\\hline
     &           &$1$\%   &$2$\%   &$3$\%  &$1$\%   &$2$\%   &$3$\%   &$1$\%   &$2$\% &$3$\%    &$1$\%   &$2$\%   &  $3$\%  \\\hline
 CBE & $-0.29$   &$-0.08$ &$-0.06$ & $-0.07$  &$-0.11$ &$-0.09$ &$-0.07$   &$-0.11$  &$-0.09$ &$-0.06$   &$0.04$  & $0.08$ & $0.12$ \\
 VBE &$2.91$     &$2.72$  &$2.74$  & $2.79$   &$2.75$  &$2.77$  &$2.79$    &$2.75$   &$2.74$  &$2.77$    &$2.60$  &$2.58 $ &$2.65$ \\\hline
 $E_{\rm gap}$ & 3.20 & 2.80 &2.80  & 2.86    & 2.86   &2.86    &2.86   &2.86    &2.83    &2.83      & 2.56     & 2.50    & 2.44  \\\hline
 \end{tabular}
}
\caption{Conduction band (CBE) and valence band (VBE) edges, both in units of eV, of pristine TNT and doped TNTs for different concentrations.
The energies are given with respect to the normal hydrogen electrode (NHE) potential; cf.\ the related discussion in section 5 of 
\cite{MMF2017}. Note that 0 eV (NHE) corresponds to $- 4.5$ eV (vacuum). In the third row, the scissors-corrected energy gap, 
$E_{\textrm{gap}} = E_{\textrm{VBE}} - E_{\textrm{CBE}}$, is presented for completeness.}
\label{table:results}
\end{table} 

Table \ref{table:results} shows that Si, Ge, and Sn mono-doping of TNTs improves the photocatalytic properties, at any concentration. 
However, the CBE value is too high compared to the reduction level of water,
hence Pb doped TNTs are useful for hydrogen production despite the fact that they have the lowest band gaps 
among the mono-dopants. The low-concentration Si and Ge doped structures show a better efficiency than for high concentration. 
In contrast, for bulk TiO$_{2}$ anatase only Ge doping improves the photocatalytic properties \cite{RLY}. We do not present
the co-doping results here because all of them have CBEs around $-2$ eV, which is higher than the reduction level of water, 
and the VBEs are higher than the oxidation level.

\section{Summary}
Density functional theory has been employed to study the structural, electronic, and op-
tical properties of cation mono- and co-doped titania nanotubes (TNTs) at different doping
concentrations. All mono-/co-dopants, except (Sn, Ge) co-doping, decrease the band gap of the TNT, similar to previous
results [40]. For mono-dopants, Pb doped TNTs have the lowest band gap at the studied concentrations (1\% to 3\%) 
due to the presence of distinct Pb states below the conduction band. The contribution of the dopant states in the 
conduction band increases as we move down the
4A group in the periodic table, i.e., from Si to Ge, Sn, and Pb. The decrease in the band
gaps of mono-doped TNTs is accompanied by shifts in the band edges towards
lower energy for Si, Ge, and Sn. The band gaps of 2\% co-doped TNTs, except for (Ge, Sn),
are smaller than those of
Si, Ge, and Sn mono-doped TNTs at any concentration. The (Pb, 2X; X = Si, Pb) co-doped TNTs have the lowest band gap of 
all mono- and co-doped TNTs. However (Ge, Sn) 2\% co-doped TNT has the largest band gap not only of all mono- and co-doped 
TNTs but also compared to the pristine nanotube. The influence of co-dopants can be
understood, to a large extent, in terms of a superposition of individual mono-dopant effects.
The study of optical properties illustrates that mono- and co-doped TNTs can absorb a
wide range of visible light, in contrast to pristine TNT. This observation, consistent with
recent experimental results, is related to the decrease of the band gap. The Si, Ge, and Sn
mono-doped TNTs at low concentration (1\%) have a high ability to produce hydrogen in the
water splitting process, their performance being clearly better than for pristine TNT. The
energetic locations of the band gap edges of Pb mono-doped and co-doped TNTs, however,
prevent their use for this application.

\acknowledgments{We thank Udo Schwingenschl\"ogl for helpful comments. Financial support from the
Deutsche Forschungsgemeinschaft (DFG, German Research Foundation -- project number 107745057 -- TRR 80)
is gratefully acknowledged.}

\end{document}